\pdfoutput=1
\documentclass[prl,twocolumn]{revtex4}%
\usepackage{amsmath}
\usepackage{amsfonts}
\usepackage{amssymb}
\usepackage{graphicx}%
\begin{document}


\title{A Perfect Metamaterial Absorber}
\author{N.~I. Landy}
\affiliation{Boston College, Department of Physics, 140 Commonwealth
Ave., Chestnut Hill, MA 02467.}
\author{S. Sajuyigbe, J.~J. Mock, D.~R. Smith}
\affiliation{Department of Electrical and Computer Engineering, Duke
University, Durham, NC 27708 USA.}
\author{W.~J. Padilla}
\affiliation{ Boston College, Department of Physics, 140
Commonwealth Ave., Chestnut Hill, MA 02467.}
\date{\today}

\begin{abstract}
We present the design for an absorbing metamaterial element with
near unity absorbance. Our structure consists of two metamaterial
resonators that couple separately to electric and magnetic fields so
as to absorb all incident radiation within a single unit cell layer.
We fabricate, characterize, and analyze a metamaterial absorber with
a slightly lower predicted absorbance of 96$\%$. This achieves a
simulated full width at half maximum (FWHM) absorbance of 4$\%$ thus
making this material ideal for imaging purposes. Unlike conventional
absorbers, our metamaterial consists solely of metallic elements.
The underlying substrate can therefore be chosen independently of
the substrate's absorptive qualities and optimized for other
parameters of interest. We detail the design and simulation process
that led to our metamaterial, and our experiments demonstrate a peak
absorbance greater than $88\%$ at 11.5 GHz.
\end{abstract}
\maketitle

The nascent field of electromagnetic metamaterials has produced
exotic effects such as negative index of
refraction~\cite{vesalago68,shelby01}, and devices such as an
electromagnetic cloak~\cite{schurig06a}. The realization of such
properties lies in the ability of metamaterials to create
independent tailored electric~\cite{pendry96} and
magnetic~\cite{pendry99} responses to incident radiation.
Electromagnetic metamaterials are also geometrically scalable which
translates into operability over a significant portion of the
electromagnetic spectrum. To date metamaterials have been
demonstrated in every technologically relevant spectral range, from
radio~\cite{wiltshire01}, microwave~\cite{smith1},
mm-Wave~\cite{ekmel}, THz~\cite{science04}, MIR~\cite{soukoulis},
NIR~\cite{brueck}, to the near optical~\cite{optical}. These
designer electromagnetic materials are an ideal platform for the
investigation of novel emergent physical phenomena while also
holding great promise for future applications.

As an effective medium \cite{drsjosab}, metamaterials can be
characterized by a complex electric permittivity
$\tilde{\epsilon}(\omega)=\epsilon_{1}+i\epsilon_{2}$ and magnetic
permeability $\tilde{\mu}(\omega)=\mu_{1}+i\mu_{2}$. Much of the
work in metamaterials has been focussed on the real part of
$\epsilon$ and $\mu$ to enable the creation of a negative refractive
material. However, the oft-overlooked loss components of the optical
constants ($\epsilon_2$ and $\mu_2$) have much potential for the
creation of exotic and useful materials as well. For example, they
can be manipulated to create a high absorber. By manipulating
electric and magnetic resonances independently, it is possible to
absorb both the incident electric and magnetic field. Additionally,
by matching $\epsilon$ and $\mu$, a metamaterial can be
impedance-matched to free space, minimizing reflectivity. In this
paper, we show that metamaterials can be fashioned to create
narrow-band perfect absorbers, with the potential to be used in
devices such as in bolometers.

\begin{figure}
[ptb]
\begin{center}
\includegraphics[width=3.25in,keepaspectratio=true
]%
{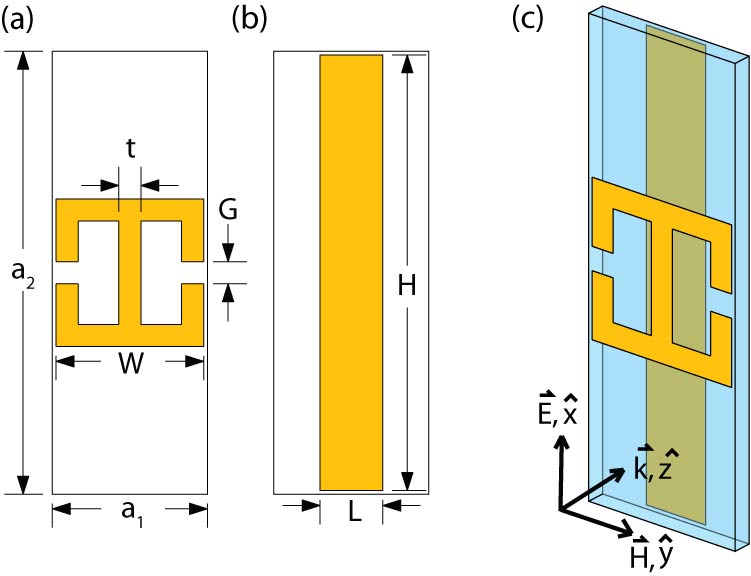}%
\caption{Electric resonator (a) and cut wire (b). Dimension
notations are listed in (a) and (b). The unit cell is shown in (c)
with axes indicating the direction of propagation of a TEM wave. }
\label{Fig1}%
\end{center}
\end{figure}

A single unit cell of the absorber consisted of two distinct
metallic elements as shown in Fig. \ref{Fig1} (a) and (b). Electric
coupling was supplied by the electric ring resonator (ERR), similar
to the design presented by Padilla \textit{et al.}~\cite{padilla07},
and is shown in Fig. \ref{Fig1} (a). This element consisted of two
standard split ring resonators connected by the inductive ring
parallel to the split-wire. We used this design instead of a
conventional split-wire design because of the limitations of
straight wire media \cite{schurig06b}, i.e. a split-wire design has
limited tunability beyond the addition of more wires per unit cell
to increase inductance~\cite{smith99}. The magnetic coupling
required a more complicated arrangement, and thus in order to couple
to the incident \textbf{H}-field, we needed flux created by
circulating charges perpendicular to the propagation vector. We
created this response by combining the center wire of the electric
resonator with a cut wire (Fig. \ref{Fig1}(b)) in a parallel plane
separated by a substrate (Fig. \ref{Fig1}(c)). This design is
similar to the so-called ``fishnet" and paired nanorod
structures\cite{brueck, optical}, in the sense that they derive a
magnetic response from driving two antiparallel currents in
conducting segments. We were then able to tune the magnetic response
by changing the geometry of the cut wire and the separation between
the cut wire and electric resonator. By manipulating the magnetic
coupling without changing the geometry of the electric resonator, we
able to decouple $\epsilon$ and $\mu$, and therefore individually
tune each resonance.

\begin{figure}
[ptb]
\begin{center}
\includegraphics[ width=3.25in,keepaspectratio=true
]%
{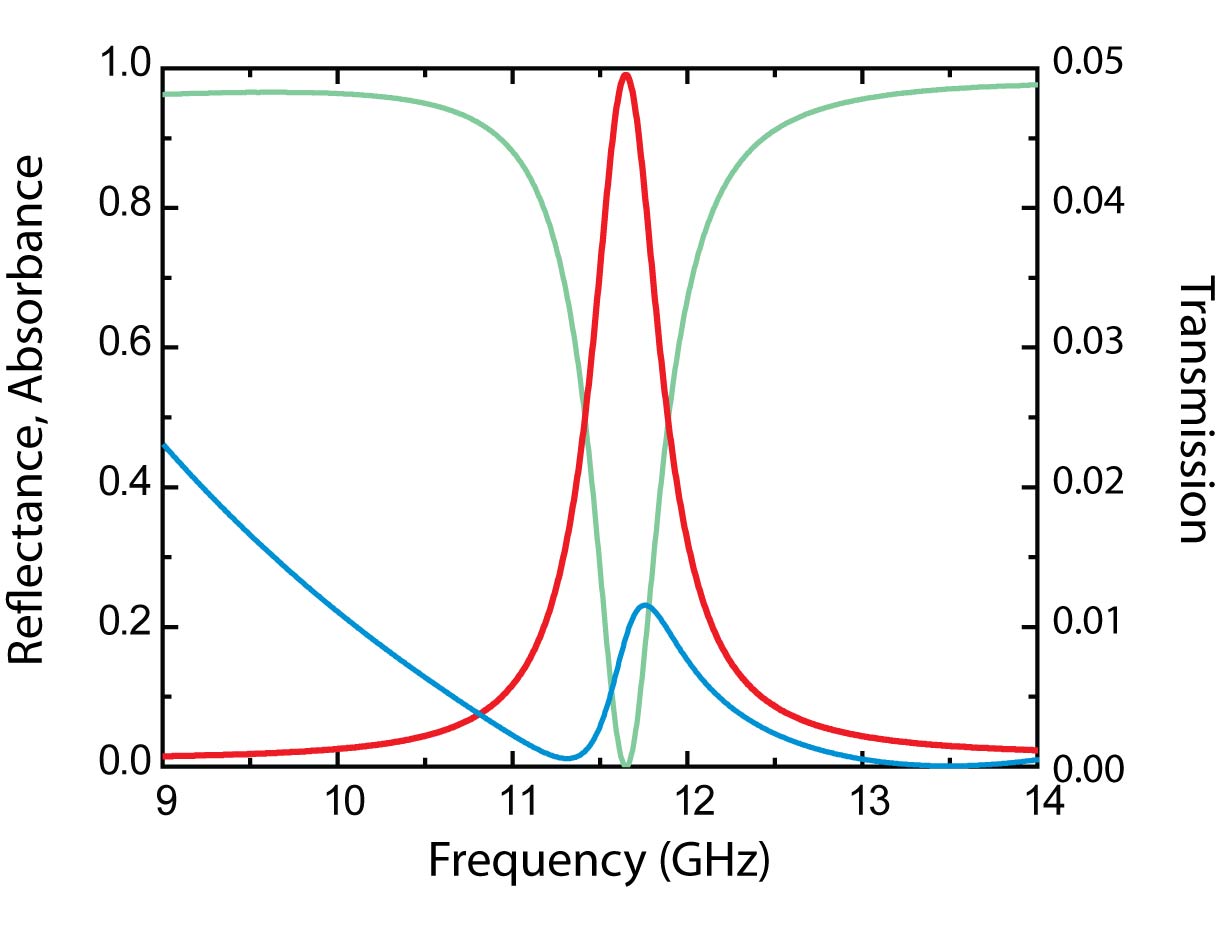}%
\caption{Simulated perfect metamaterial absorber. Reflectance (green
curve) and absorbance (red curve) are plotted from zero to 100$\%$
(left axis). The transmission is plotted on the right axis as the
blue curve on a scale from zero to 5$\%$. }
\label{Fig2a}%
\end{center}
\end{figure}

We performed computer simulations of an ideal -- but realizable --
metamaterial absorber using the commercial finite-difference time
domain (FDTD) solver Microwave Studio by CST.\cite{cst} The program
simulated a single unit cell as shown in Fig. \ref{Fig1} (c) with
appropriate boundary conditions, i.e. perfect electric
($\hat{y}\hat{z}$-plane) and perfect magnetic
($\hat{x}\hat{z}$-plane). Waveguide ports on the other boundaries
simulated a TEM plane wave propagating through the medium.
Simulation produced the complex frequency dependent S-parameters,
$\widetilde{S}_{11}$ and $\widetilde{S}_{21}$, where
$T(\omega)=|S_{21}|^2$ and $R(\omega)=|S_{11}|^2$ are the
transmission and reflectance, respectively. From the S-parameter
data we inverted the fresnel equations to extract the complex
optical constants\cite{smith02}. We then examined the behavior of
the surface current density, magnetic and electric fields at
$\omega_0$ to verify that we were coupling to the correct resonant
mode of each metamaterial element. By tuning $\omega_0$ of the
$\epsilon$ and $\mu$ resonances (not shown), due to the two
metamaterial elements, we achieved $\epsilon=\mu$ and thus an
impedance near the free space value. While $100\%$ asborbance is
theoretically possible, it can only occur when the metamaterial
layer is impedance-matched to free space such that the reflectance
is zero.\cite{joannopoulos02} One may then minimize transmission
such that the addition of multiple layers ensures
$T(\omega)\rightarrow0$. The simulated metamaterial had the
dimensions, in millimeters, of: a$_1$=4.2, a$_2$=12, W=3.9, G=0.606,
t=0.6, L=1.7, H=11.8, and the metamaterials elements were separated
by 0.65 in the $\hat{z}$ direction. In Fig. \ref{Fig2a} we show the
simulation results for the metamaterial perfect absorber.
Reflectance and absorbance are plotted from zero to 100$\%$ (left
axis) and the transmission is plotted on the right axis from zero to
5$\%$. The reflectance is large $\sim97\%$ near the bounds of the
plot, 9GHz and 14GHz, but there is a minimum of 0.01$\%$ at
$\omega_0\equiv11.65$GHz. The simulated transmission also undergoes
a minimum near $\omega_0$ and yields a value of $\sim0.9\%$. Thus we
achieve a best simulated absorbance
$A(\omega)=1-T(\omega)-R(\omega)$ slightly less than unity
$A(\omega)=99\%$ with a FWHM of 4$\%$ compared to $\omega_0$.

\begin{figure}
[ptb]
\begin{center}
\includegraphics[ width=3.25in,keepaspectratio=true
]%
{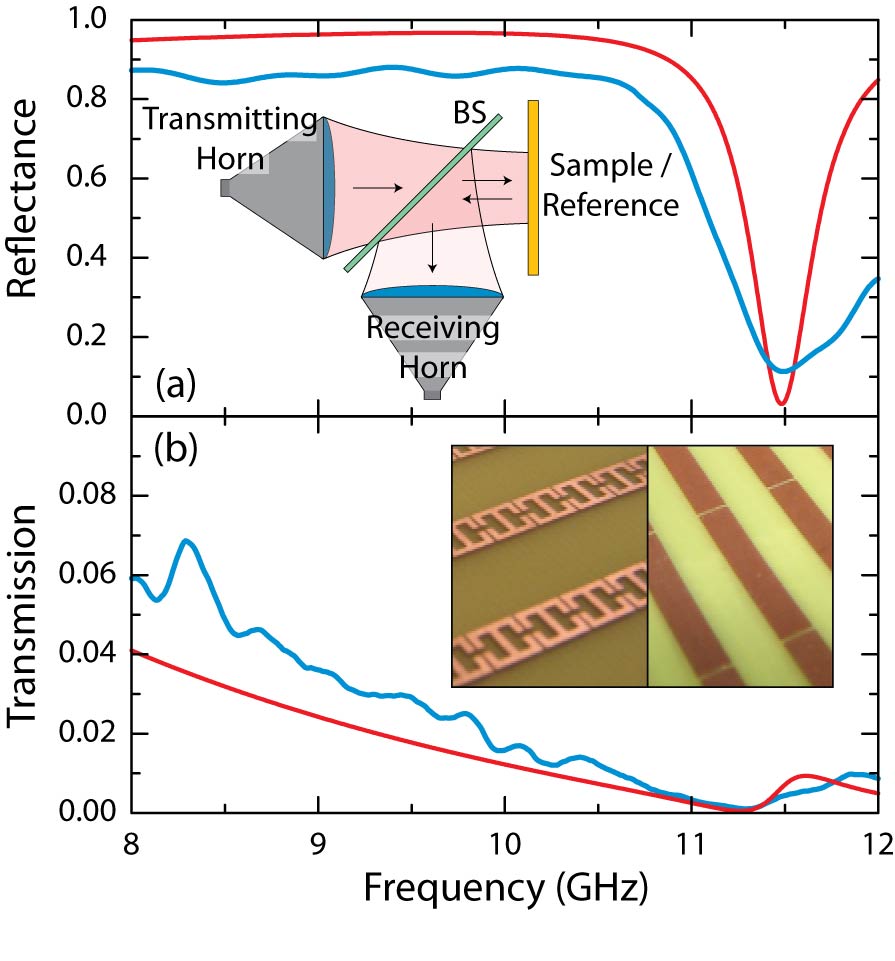}%
\caption{Simulated (red curves) and measured (blue curves)
T($\omega$) and R($\omega$) for the microwave absorber. (a) displays
the simulated and measured reflectance from zero to 100$\%$. A
schematic of the reflection experiment is shown as the inset to (a)
where BS is the beam splitter. The simulated and measured
transmission, shown in (b), are displayed from zero to 10$\%$. Both
reach a minimum of approximately $0.6\%$ near 11.5 GHz. Inset to (b)
shows photographs of the individual components which make up the
metamaterial absorber. The sample consists of the electric ring
resonator (left) and the split wire (right) and is joined with an
FR4 spacer of 0.72mm thickness.}
\label{Fig3a}%
\end{center}
\end{figure}

While our simulated absorber achieves narrow-band high absorbance,
we were restricted by the minimum line widths available to us (250
$\mu$m) and other various fabrication tolerances. These limitations
were incorporated into the design process and we fabricated a
metamaterial which deviated slightly from the ideal absorber and had
the dimensions, in millimeters, of: a$_1$=4.2, a$_2$=12, W=4, G=0.6,
t=0.6, L=1.7, H=11.8. Each metallization was fabricated on a FR4
substrate with a thickness of 0.2mm. Metamaterials were fabricated
using standard optical lithography. Photosensitized half-ounce
copper-clad FR-4 circuit board formed the substrate material with an
17 $\mu$m copper thickness. A mask was designed and printed in high
resolution on a transparency. The photosensitized copper clad FR-4
was covered by the mask and exposed to ultraviolet light then
developed in a sodium carbonate solution and etched in ferric
chloride. Finally, the board was placed in a stripper solution to
remove the remaining photoresist. Both the ERR layer (Fig.
\ref{Fig1}(a)) and the cut-wire layer (Fig. \ref{Fig1}(b)) were
fabricated in such a fashion, see inset to Fig. \ref{Fig3a} (b).
These boards where then sandwiched (using an adhesive with 0.06mm
thickness) about another 0.2mm thick FR4 blank substrate to obtain
the correct spacing. The end results was that the metamaterials
elements were separated by 0.72mm in the $\hat{z}$ direction. This
design allowed us to couple to both magnetic and electric resonances
without using out-of-plane elements typical with ``wine-crate"
designs,\cite{shelby01} thus greatly simplifying construction of the
absorber.

We experimentally verified the behavior of the absorber by measuring
the complex S-parameters of a large planar array of pixels (outer
dimensions of 15 $\times$ 15 cm). We used an Agilent vector network
analyzer that produced microwaves in the range of 8-12 GHz. One
microwave horn (Rozendal and Associates) focused the GHz beam on the
sample and another horn served as a detector. Both horns coupled to
linearly polarized light and had parallel polarization directions.
To measure transmission, we set the horns in a normal incidence
confocal configuration. We performed reflectance measurements at
normal incidence. However, in order to eliminate large voltage
standing wave ratio (VSWR) which would overwhelm the data, we used a
beam splitter configuration, shown as the inset to Fig. \ref{Fig3a}
(a). Radiation first traveled from the transmitting horn through the
beam splitter, reflected from the sample and returned to the beam
splitter where it was reflected at 45$\%$ to the receiving horn. We
set the normalization and phase in the transmission configuration by
simply removing the sample. For reflection, we replaced the sample
with a perfect reflector for normalized measurements. Transmission
characterization techniques utilized here have been described
elsewhere in detail\cite{starr04}.

\begin{figure}
[ptb]
\begin{center}
\includegraphics[ width=3.25in,keepaspectratio=true
]%
{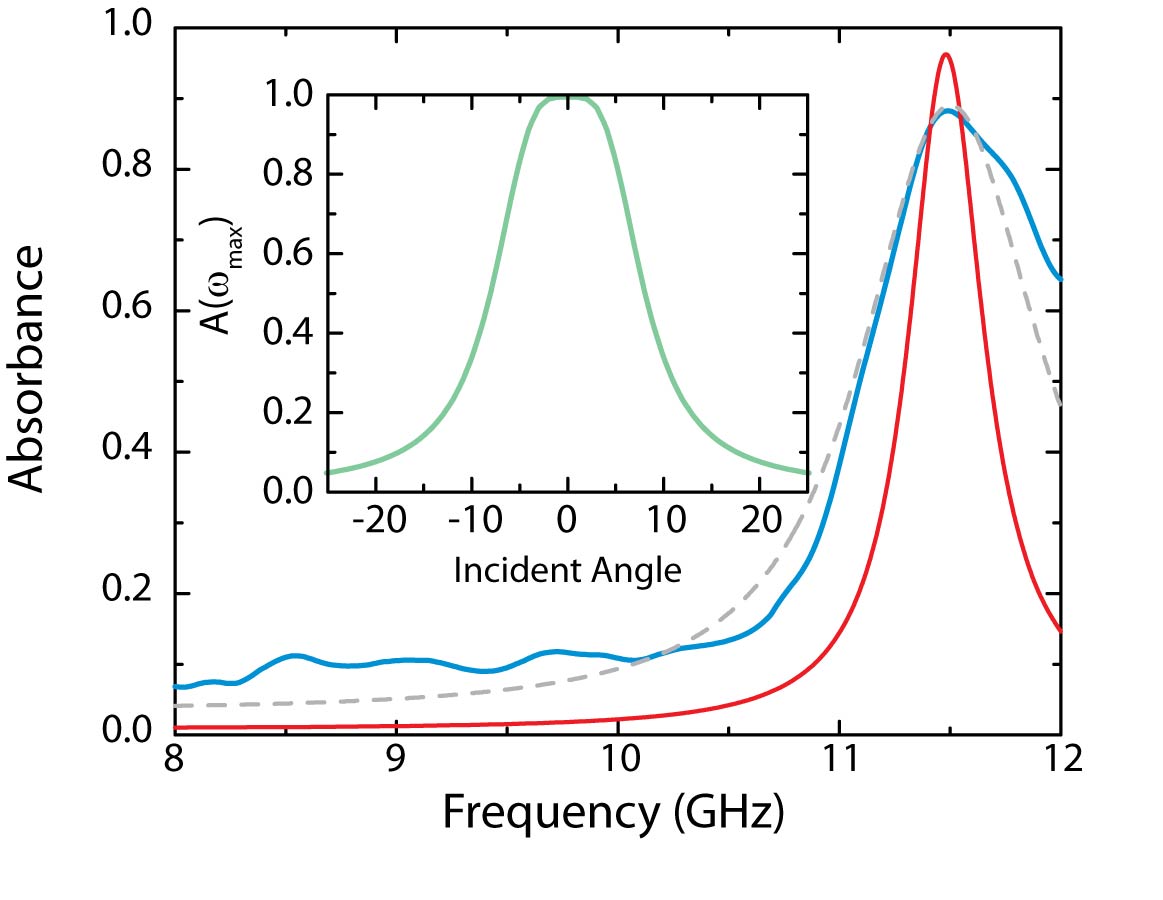}%
\caption{Main panel shows the simulated (red) and measured (blue)
absorbance curves. Significant broadening in the reflection curve
contributes to the deviation of the experimental absorbance curve
from simulation. The dashed gray absorbance curve is a Gaussian
weighted average which approximates assembly error of the
metamaterials absorber, described in the text. Inset
shows the simulated angular dependence of the absorbance at at
$\omega_{max}$.}
\label{Fig4}%
\end{center}
\end{figure}

Computer simulations were performed for this metamaterial structure
and are presented for comparison to experimental data. The simulated
$R(\omega$) is shown as the red curve in Fig. \ref{Fig3a} (a). At
low frequencies the reflectivity is high and yields values near
95$\%$. A large feature occurs in $R(\omega)$ and has a minimum of
$\sim$~3$\%$ near 11.5 GHz. Likewise the simulated transmissivity,
shown as the red curve in Fig. \ref{Fig3a} (b), has 4$\%$
$T(\omega)$ at low frequencies and then undergoes a minimum near
11.25 GHz. The simulated absorbance of our metamaterial unit cell
thus has a maximum near where both $R(\omega)$ and $T(\omega)$ have
their minima. We plot $A(\omega)$ as the red curve in Fig.
\ref{Fig4} from 8 to 12 GHz. The simulated $A(\omega)$ peaks at
96$\%$ at 11.48 GHz and has a FWHM of 4$\%$ with respect to its
center frequency.

By performing simulations for transmissivity and reflectivity, we
are able to account for the form of the experimental data. The
measured $R(\omega)$ and $T(\omega)$ are plotted as the blue curves
in Fig. \ref{Fig3a} (a) and (b) respectively. We find excellent
agreement between simulated and measured T$(\omega)$ over all
frequencies characterized. The measured reflectance is consistently
$8\%$ lower than the simulated $R(\omega)$ at low frequencies. Both
the simulated and experimental R$(\omega)$ reach a minimum at
approximately 11.5 GHz, but experimentally the minimum is $11\%$, as
opposed to the simulated value of $3\%$. There is also significant
broadening in the experimental reflectivity curve with respect to
simulation. This is most likely due to tolerances in the fabrication
and assembly.

\begin{figure}
[ptb]
\begin{center}
\includegraphics[ width=3.25in,keepaspectratio=true
]%
{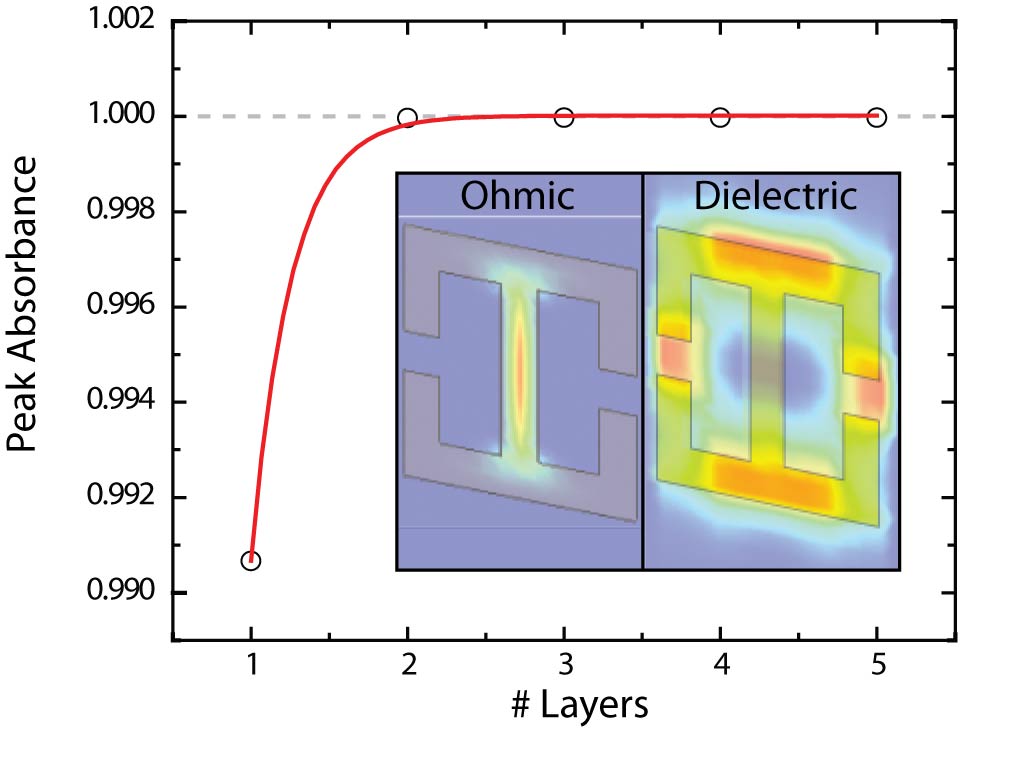}%
\caption{Simulated absorbance with increasing metamaterial layers,
red curve is a guide to the eye. Insets show the simulated losses at
resonance. The Ohmic loss (surface) is shown in the left inset panel
and the dielectric (volume) loss is on the right.}
\label{Fig5}%
\end{center}
\end{figure}

The experimental absorbance calculated from $T(\omega)$ and
$R(\omega)$ is shown as the blue curve in Fig. \ref{Fig4}. The
simulated and experimental curves have maximum absorbance at the
same frequency, $\omega_{max}$=11.5 GHz. However, the simulated
curve reaches a maximum of 96$\%$ and the experimental curve
only reaches 88$\%$. The difference between the simulated and
measured curves is due primarily to deviations in $R(\omega)$.
Better fabrication and assembly tolerances will permit absorbances
and bandwidths similar to that shown as the red curve in Fig.
\ref{Fig4}. We performed computer simulations to verify the origin
of the broadening of A($\omega$). From a theoretical viewpoint it is
expected that the frequency location of the absorbance peak depends
significantly upon the separation between the electric ring
resonator and the cut-wire, since this distance determines the
frequency location of the magnetic resonance. Computer simulations verified this
conjecture. We simulated various spacings centered about the
designed value of 0.72mm, and computed a Gaussian weighted average
$A(\omega)$ with a standard deviation of $\sigma=$20 $\mu$m. The absorbance
calculated in this manner agrees very well with the experimental
curve and is plotted as the grey dashed curve in the main panel of Fig.
\ref{Fig4}. Thus assembly errors in the metamaterial spacing of only
$5\%$ can easily account for the disagreement between the simulation
and experiment shown in Fig. \ref{Fig4} as the red and blue curves,
respectively.

An investigation into the effect of adding multiple metamaterial
absorbing layers is shown in the main panel of Fig. \ref{Fig5}. As
can be observed the absorbance rises sharply with additional layers
and is asymptotic to unity, within computational noise. Two layers
of the metamaterial absorber achieve a value of $\sim$99.9972\%. At
this thickness the entire metamaterial is only $\sim\lambda_0/2$ at
resonance. For the structure shown in Fig. \ref{Fig1}, A($\omega_0$)
can never reach unity since the metamaterial is not perfectly
matched to free space. However, it is important to note that
although an impedance matched structure may yield zero
R($\omega_0$), this does not necessarily mean the transmission will
be small. In this study we are interested in achieving
R($\omega_0$)=T($\omega_0$)=0 in a single unit cell in the
propagation direction. Thus our metamaterial structure was optimized
to maximize the absorbance with the restriction of minimizing the
thickness. If this constraint is relaxed, however, an impedance
matched condition can be obtained and with multiple layers a perfect
absorbance can be achieved.

We further investigated the origin of the loss in our metamaterial
structure, as indicated in the inset to Fig. \ref{Fig5}. The Ohmic
``surface" loss originates mainly from the center conducting region
of the electric ring resonator, and dielectric losses primarily
occur in-between the two metamaterial elements where the electric
field is large. The major component responsible for the absorbance
in our metamaterial structure is the dielectric loss and simulation
indicates this is an order of magnitude greater than Ohmic losses.
This is consistent with previous studies of frequency selective
surfaces (FSSs) where it was found that metallic absorption was
relatively insignificant in comparison to dielectric
losses.\cite{munk}

We now discuss the potential application of a perfect metamaterial
absorber as a bolometer. An ideal bolometric detector is one in
which all photons that fall upon its surface are absorbed, converted
to heat, and thus sensed.\cite{richards94} A prominent feature of
commercially available bolometers is their extreme bandwidth making
them ideal candidates as detectors, especially in the far infrared
frequency regime where other efficient detection methods are largely
unavailable. However, bolometers which achieve a narrowband response
also have significant application as focal plane array detectors for
imaging. The metamaterial design presented here achieves simulated
absorbtivities near unity making them ideal candidates for
bolometric pixel elements. The natural narrowband metamaterial
resonance is a salient feature for focal plane array detectors as it
is naturally apodizing and operates at room temperature. Since the
elements which constitute our bolometer are sub-wavelength,
metamaterials can inherently image at the diffraction limit.
Further, here we have used only a \textit{single unit cell} in the
propagation direction, (with a thickness significantly smaller than
the wavelength $\lambda_0/35$), yet achieved an experimental
absorbance of ~88$\%$. By adding multiple layers one can achieve
narrowband absorptivities of unity. Although the design is planar
and A($\omega$) should fall off relatively rapidly, we investigated
the angular dependence (inset to Fig. \ref{Fig4}) and found the
metamaterial still achieved 50$\%$ absorbance at an incident
full-angle of 16$^{\circ}$. Finally, the scalability of
metamaterials permits the usage of these perfect absorbers at other
wavelengths of interest such as the possibility of room temperature
high resolution imaging at mm-Wave and THz frequencies. This is
particularly interesting owing to the many possible applications. We
have designed similar metamaterials (not shown) as in Fig.
\ref{Fig1} that achieved comparable $A(\omega)$ to that shown in
Fig. \ref{Fig2a} operating at 94 GHz and 1 THz.

While the creation of a metamaterial absorber is novel, the
technology to create highly absorptive materials in the microwave
frequency range is well-established. The material ``chiroshield" is
capable of very high absorption, and reduces backscatter by 15 to 25
dB \cite{chiral}. Chiroshield, however, has geometries on the order
of one wavelength, which is 35 times larger than our design at a
given center frequency. Some ferrites show reflection loss of -30 dB
\cite{singh00}. These ferrites, however, lack the inherent and
precise tunability of metamaterials. Other work has been done on the
application of frequency-selective materials to bolometric devices.
These bolometric designs, however, require cryogenic temperatures to
operate and only achieve absorbances of about $50\%$\cite{perera06}.
At optical wavelengths, metal colloides are known to have large
absorptive properties,\cite{lamb80} due to the geometric specific
surface modes\cite{pendry94}. This suggests and interesting possible
application of artificial plasmonic wire media\cite{pendry96} shaped
into particular geometries to function as perfect
absorbers\cite{smith99} operating at microwave frequencies.

In conclusion, we have demonstrated that metamaterials can be highly
absorptive over a narrow frequency range. This stems from the
ability to design metamaterial elements which can individually
absorb the electric and magnetic components of an incident
electromagnetic wave. In contrast previous experimental results at
THz frequencies using a single type of metamaterial element
(electric) yielded values of only
$A(\omega)$=$\sim20\%$.\cite{johara} The resonant frequency of the
metamaterial elements may be tuned throughout some range of
frequencies \cite{josab,houtong08} thus enabling hyperspectral
imaging. A further benefit afforded by metamaterials is the ability
to construct a single unit cell with $\mu(\omega)=\epsilon(\omega)$
over an extended frequency range. Thus this unit cell can achieve
zero reflectance since it can have an impedance equal to the free
space value $Z=\sqrt{\mu/\epsilon}=1$. This is similar to the well
known theoretical construct, the Perfectly Matched Layer (PML)
Absorbing Boundary Condition (ABC),\cite{pml} which splits waves
incident upon a boundary into electric and magnetic components to
obtain near perfect absorbtion. The PML, however, requires gain and
additionally is on the order of $\lambda_0/2$ in
thickness\cite{compbook}.

The design presented here can still be improved. By incorporating a
substrate with a highly consistent dielectric constant, we will be
able to optimize the design at the correct resonant frequency.
Similarly, tighter fabrication tolerances would allow us to bring
the impedance closer to unity and therefore closer to $100\%$
absorption. More progress could also be made on the design itself.
Currently, the absorber is polarization sensitive, which is not
ideal for some applications. Additionally, the assembly is
complicated due the restrictions placed on the direction of the
incident wave to couple to the magnetic resonator. This resulted in
variations between unit cells and less distinct absorbance peak.

NIL and WJP acknowledge support from the Los Alamos National
Laboratory LDRD program, DRS acknowledges support MDA Contract
W9113M-07-C-0078, and DRS, SS, and JJM acknowledge support from a
Multiple University Research Initiative sponsored by the Air Force
Office of Scientific Research Contract No. FA9550-06-1-0279.

\end{document}